\documentclass[12pt]{article}

\RequirePackage{etex}

\usepackage[margin=1in]{geometry}  	
\usepackage{amsfonts}
\usepackage{amsmath} 
\usepackage{amssymb}
\usepackage{centernot}
\usepackage{booktabs}
\usepackage{multirow}
\usepackage{algorithm}
\usepackage{algorithmicx}

\usepackage[onehalfspacing]{setspace} 
\renewcommand{\arraystretch}{0.6} 

\usepackage{float}
\usepackage{caption}
\usepackage{graphicx}
\usepackage{adjustbox}
\usepackage{lscape}

\usepackage{adjustbox}

\usepackage{xcolor}

\usepackage{relsize}

\usepackage{sectsty}
\sectionfont{\normalsize\MakeUppercase}
\subsectionfont{\normalsize}
\subsubsectionfont{\itshape}

\usepackage{bbm}
\usepackage{subfig}

\definecolor{forestgreen}{RGB}{34,139,34}
\usepackage{hyperref}
\hypersetup{
    colorlinks=true,
    linkcolor=magenta,
    filecolor=magenta, 
    citecolor=forestgreen,      
    urlcolor=blue
}

\usepackage{authblk}

\usepackage{amsthm}
\newtheorem{theorem}{Theorem}
\newtheorem{proposition}[theorem]{Proposition}

\usepackage{xpatch}
\makeatletter
\xpatchcmd{\proof}{\@addpunct{.}}{\@addpunct{:}}{}{}
\makeatother

\makeatletter
\def\@hangfrom#1{\setbox\@tempboxa\hbox{{#1}}%
      \hangindent 0pt
      \noindent\box\@tempboxa}
\makeatother

\makeatletter
\newcommand{\vast}{\bBigg@{3}}
\newcommand{\Vast}{\bBigg@{4}}
\makeatother

\makeatletter
\newcommand*{\indep}{%
  \mathbin{%
    \mathpalette{\@indep}{}%
  }%
}
\newcommand*{\nindep}{%
  \mathbin{
    \mathpalette{\@indep}{\not}
  }%
}
\newcommand*{\@indep}[2]{%
  \sbox0{$#1\perp\m@th$}
  \sbox2{$#1=$}
  \sbox4{$#1\vcenter{}$}
  \rlap{\copy0}
  \dimen@=\dimexpr\ht2-\ht4-.2pt\relax
  \kern\dimen@
  {#2}%
  \kern\dimen@
  \copy0 
} 
\makeatother

\DeclareMathOperator{\E}{\textnormal{\mbox{E}}}

\usepackage[utf8]{inputenc}
\DeclareUnicodeCharacter{200E}{}

\usepackage{tikz}
\usetikzlibrary{positioning,shapes.geometric}

\usepackage{setspace}
\usepackage{cite}

\usepackage[stable]{footmisc}

\usepackage{rotating}

\makeatletter
\def\@seccntformat#1{\@ifundefined{#1@cntformat}%
   {\csname the#1\endcsname\quad}  
   {\csname #1@cntformat\endcsname}
}
\let\oldappendix\appendix 
\renewcommand\appendix{%
    \oldappendix
    \newcommand{\section@cntformat}{\appendixname~\thesection\quad}
}
\makeatother

\usepackage[absolute,showboxes]{textpos}

\TPMargin{5pt}

\usepackage{thmtools, thm-restate}

\usepackage{datetime}

\begin{document}

\title{\textbf{Transporting a prediction model for use in a new target population} \vspace*{0.3in} }

\author[1]{Jon A. Steingrimsson\footnote{Address for correspondence: Dr. Jon A. Steingrimsson; Department of Biostatistics, School of Public Health, Brown University; 121 South Main Street, Providence, RI 02903;
email: \texttt{jon\_steingrimsson@brown.edu}.}}
\author[1]{Constantine Gatsonis}
\author[2]{Issa J. Dahabreh}

\affil[1]{Department of Biostatistics, School of Public Health, Brown University, Providence, RI}
\affil[2]{Departments of Epidemiology and Biostatistics, Harvard T.H. Chan School of Public Health, Boston, MA}

\maketitle{}

\thispagestyle{empty}

\newpage
\thispagestyle{empty}

\vspace*{1in}

\begin{abstract}
\noindent
\linespread{1.7}\selectfont We consider methods for transporting a prediction model and assessing its performance for use in a new target population, when outcome and covariate data for model development are available from a simple random sample from the source population, but only covariate data are available from a simple random sample from the target population. We discuss how to tailor the prediction model for use in the target population, how to assess model performance (e.g.,~by estimating the target population mean squared error), and how to perform model and tuning parameter selection. We provide identifiability results for measures of performance in the target population for a potentially misspecified prediction model under a sampling design where the source and the target population samples are obtained separately. We also introduce the concept of prediction error modifiers that can be used to reason about tailoring measures of model performance to the target population. We illustrate the methods using simulated data.

\vspace{0.3in}

\end{abstract}

\clearpage 
\section*{Introduction}
\setcounter{page}{1}

Users of prediction models typically want to obtain predictions in a specific target population. For example, a healthcare system may want to deploy a clinical risk prediction model \cite{steyerberg2019clinical} to identify individuals at high risk for adverse outcomes among all patients receiving care. Prediction models are often built using data from source populations represented in prospective epidemiological cohorts, confirmatory randomized trials \cite{pajouheshnia2019and}, or administrative databases \cite{goldstein2017opportunities}. In most cases, the data from the source population that are used for developing the prediction model cannot be treated as a random sample from the target population where the model will be deployed because the two populations have different data distributions. Consequently, a model developed using the data from the source population may not be applicable to the target population and model performance estimated using data from the source population may not reflect performance in the target population. 

Consider a setup where outcome and covariate data are available from a sample of the source population and only covariate data are available from a sample of the target population. For example, covariate data from the target population may be obtained from administrative databases, but outcome data may be unavailable (e.g.,~when outcome ascertainment requires specialized assessments) or insufficient (e.g.,~when the number of outcome events is small due to incomplete followup). In this setup, developing and assessing the performance of a prediction model for the target population is not possible using standard methods because of the complete lack of outcome data from the target population; using data from the source population can be an attractive alternative. Yet, as noted above, directly applying a prediction model developed in data from the source population to the target population, or treating model performance measures (e.g.,~mean squared prediction error) estimated in the source data as reflective of performance in the target population may be inappropriate when the two populations have different data distributions. Thus, investigators are faced with two transportability tasks: (1) tailoring a prediction model for use in a target population when relying on outcome data from the source population; and (2) assessing the performance of the model in that target population.

These two transportability tasks have received attention in the computer science literature on covariate shift and domain adaptation \cite{bickel2007discriminative, sugiyama2007covariate, pan2010domain, cao2011distance, sugiyama2012machine, kouw2018introduction, chen2019tailoring, ishii2020partially, datta2020regularized}. In epidemiology, however, the transportability of prediction models has been treated heuristically and commonly used methods do not have well-understood statistical behavior. The related problem of transporting inferences about treatment effects to a target population has received more attention \cite{cole2010generalizing, rudolph2017robust, dahabreh2020extending, dahabreh2019generalizing}, but there are important differences between transportability of treatment effects and prediction models in terms of the parameters being estimated and the methods used for estimation. 

Here, we examine the conditions that allow transporting prediction models from the source population to the target population. We discuss the implications of these conditions both for tailoring the models for use in the target population and for assessing model performance in that context. We show that many popular measures of model performance can be identified and estimated using covariate and outcome data from the source population and just covariate data from the target population under both nested and non-nested sampling designs, without the strong assumption that the prediction model is correctly specified. We discuss the relevance of our results when using modern model-building approaches such as cross-validation-based model selection. We introduce the concept of prediction error modifiers, which is useful for reasoning about transportability of measures of model performance to the target population. Last, we illustrate the methods using simulated data.

\section*{Sampling design and identifiability conditions}
\label{sec:DS}

Let $Y$ be the outcome of interest and $X$ a covariate vector. We assume that outcome and covariate information is obtained from a simple random sample from the source population $\{(X_i,Y_i): i = 1, \ldots, n_{\text{\tiny source}}\}$. Furthermore, covariate information is obtained from a simple random sample from the target population, $\{X_i: i = 1, \ldots, n_{\text{\tiny target}}\}$; no outcome information is available from the target population. This ``non-nested'' sampling design \cite{dahabreh2019extendingA, dahabreh2019study}, where the samples from the target and source population are obtained separately, is the one most commonly used in studies examining the performance of a prediction model in a new target population. For that reason, we will present results for non-nested designs in some detail, before considering nested designs, where the source population is a subset of a larger population that represents the target population.

Let $S$ be an indicator for the population from which data are obtained, with $S=1$ for the source population and $S=0$ for the target population, and denote $n = n_{\text{\tiny source}} + n_{\text{\tiny target}}$ as the sample size of the composite dataset consisting of the data from the source and target population samples. This composite dataset is randomly split into a training set and a test set. The training set is used to build a prediction model for the expectation of the outcome conditional on covariates in the source population, $\E[Y|X, S=1]$, and then, the test set is used to evaluate model performance. We use $g_{ \beta}(X)$ to denote the posited parametric model, indexed by the parameter $\beta$, and $g_{ \widehat \beta}(X)$ to denote the ``fitted'' model with estimated parameter $\widehat \beta$. We use $f(\cdot)$ to generically denote densities.

We assume the following identifiability conditions:
\begin{enumerate}
\item[A1.] \textit{Conditional independence of the outcome $Y$ and the data source $S$}. For every $x$ with positive density in the target population, $f(X = x,S=0) > 0$, 
\[
f(Y|X=x, S=1) = f(Y|X=x, S=0).
\]
Informally, this condition means that the relationship between $Y$ and $X$ is the same in the source population and the target population and it implies that the conditional expectation of $Y$ given $X$ is the same in the two populations, $\E[Y|X,S=1] = \E[Y|X, S=0]$. 
\item[A2.] \textit{Positivity}. For every $x$ such that $f(X=x,S=0) \neq 0$, $\Pr[S=1|X=x] > 0$. Informally, this condition means that every covariate pattern in the target population can occur in the source data, as sample size goes to infinity. 
\end{enumerate}

Next, we discuss how, under assumptions A1 and A2, the prediction model can be tailored for use in the target population and how we can assess model performance in the target population.

\section*{Tailoring the model to the target population}\label{sec:cond}

Recall that $g_\beta(X)$ is a model for $\E[Y|X,S=1]$. Suppose that the parameter $\beta$ takes values in the space $\mathcal{B}$. We say that the model is correctly specified if there exists a $\beta_0 \in \mathcal{B}$ such that $g_{\beta_0}(X) = \E[Y|X,S=1]$ \cite{wooldridge2010econometric}. Tailoring the fitted model $g_{\widehat \beta}(X)$ for use in the target population depends on whether the posited model $g_\beta(X)$ is correctly specified. 

\paragraph{Correctly specified model:} Suppose that the model $g_\beta(X)$ is correctly specified and thus we can construct a model-based estimator $g_{\widehat \beta}(X)$ that consistently estimates $\E[Y|X,S=1]$. Under condition A1, a consistent estimator for $\E[Y|X,S=1]$ is also consistent for $\E[Y|X,S=0]$ (because the two expectations are equal when condition A1 holds). Moreover, when the model for the conditional expectation is parametric (as we have assumed up to now) and the parameter $\beta$ is estimated using maximum likelihood methods, then the unweighted maximum likelihood estimator $\widehat \beta$ estimated using only the source data training set is optimal in terms of having the smallest asymptotic variance \cite{imbens1996efficient, shimodaira2000improving}. 

\paragraph{Missspecified model:} Now, suppose, as is more likely to be the case, that the model $g_\beta(X)$ is misspecified. In that case, theoretical work on the behavior of weighted maximum likelihood estimators for $\beta$ under covariate shift \cite{shimodaira2000improving} shows that the maximum likelihood estimator estimated using  only source population data is no longer optimal, in the sense of  minimizing the Kullback-Leibler divergence between the estimated and true conditional density of the outcome given covariates. Instead, the Kullback-Leibler divergence is minimized by using a 
weighted maximum likelihood estimator with weights set equal to the ratio of the densities in the target and source populations, that is, $f(X|S = 0) / f(X | S = 1)$. 

In applied work, the density ratio is typically unknown and needs to be estimated using the data, but direct estimation of density ratios is challenging, particularly when $X$ is high-dimensional \cite{sugiyama2012density}. Instead, we can use the fact that the density ratio is, up to a proportionality constant, equal to the inverse of the odds of being from the source population, $$ \dfrac{f(X|S = 0)}{f(X | S = 1)} \propto \dfrac{\Pr[ S = 0 |X]}{\Pr[S = 1|X]},$$ to replace density ratio weights with inverse-odds weights and obtain an optimal estimator of the model, tailored for use in the target population. The inverse-odds weights can be obtained by estimating the probability of an observation being from the source population conditional on covariates -- a task for which many practical methods are available for high-dimensional data \cite{hastie2009elements}. A reasonable approach for tailoring a potentially misspecified prediction model for use in the target population could proceed in three steps. Fist, estimate the probability of being from the source population, using training data from the source population and target population. Second, use the estimated probabilities to construct inverse-odds of participation weights for observations in the training set from the source population. Third, apply the weights from the second step to estimate the prediction model using all observations in the training set from the source population.

One difficulty with the above procedure is that, in non-nested designs, the sample from the source population and the sample from the target population are obtained separately, with sampling fractions from the corresponding underlying populations that are unknown by the investigators and unlikely to be equal. When that is the case, the probabilities $\Pr[S=0|X]$ and $\Pr[S=1|X]$ in the inverse-odds weights are not identifiable from the observed data \cite{dahabreh2019study, dahabreh2020benchmarking} (i.e.,~cannot be estimated using the observed data). Although the inverse-odds weights are not identifiable, in Appendix \ref{app:ID2} we show that, up to an unknown proportionality constant, they are equal to the  inverse-odds of participation weights \emph{in the training set},
\begin{equation}
\label{OR-ID}
\dfrac{\Pr[ S = 0 |X]}{\Pr[S = 1|X]} \propto \dfrac{\Pr[ S = 0 |X,D_{\text{\tiny train}}=1]}{\Pr[S = 1|X,D_{\text{\tiny train}}=1]},
\end{equation}
where $D_{\text{\tiny train}}$ is an indicator if data from an observation is in the training set and used to estimate the inverse-odds weights. It follows that we can use inverse-odds weights estimated in the training set, when estimating $\beta$ with the weighted maximum likelihood estimator.


\section*{Assessing model performance in the target population} \label{sec:Mar}

We now turn our attention to assessing model performance in the target population. For concreteness, we focus on model assessment using the squared error loss function and on identifying and estimating its expectation, that is, the mean squared error (MSE), in the target population. The squared error loss $(Y - g_{\widehat \beta}(X))^2$ quantifies the discrepancy between the (observable) outcome $Y$ and the model-derived prediction $g_{\widehat \beta}(X)$ in terms of the square of their difference. The MSE in the target population is defined as $$\psi_{\widehat \beta} = \E[(Y - g_{\widehat \beta}(X))^2|S=0].$$ In the main text of this paper, we focus on the MSE because it is a commonly used measure of model performance. Our results, however, readily extend to other measures of performance. In Appendix \ref{app:ID}, we provide identifiability results for general loss function-based measures of model performance. 

\subsection*{Prediction error modifiers}

To help explain why model performance measures need to be tailored for use in the target population, we introduce the term ``prediction error modifier'' to describe a covariate that, for a given prediction model, is associated with prediction error as assessed with some specific measure of model performance. 
Slightly more formally and using the squared error loss as an example, we say that the random variable $Z$ is a prediction error modifier, for the model $g_{\widehat \beta}(X)$, with respect to MSE in the source population, if the conditional expectation $\E[(Y - g_{\widehat \beta}(X))^2|Z = z , S=1]$ varies as a function of $z$. Several parametric or non-parametric methods are available to examine whether $\E[(Y - g_{\widehat \beta}(X))^2|Z, S=1]$ is a constant \cite{luedtke2019omnibus}. The prediction error modifier $Z$ can contain all the covariates in $X$ or only a subset of them. When the distribution of prediction error modifiers differs between the source and target populations, measures of model performance estimated using data from the source population are unlikely to be applicable in the target population, in the sense that the performance of the model in the source data may be very different (either better or worse) compared to performance of the same model in the target population. Large differences in performance measures between the source and target population can occur even if the true outcome model in the two populations is the same (i.e.,~even if condition A1 holds) because most common measures of model performance average (marginalize) prediction errors over the data distribution of the target population, and the covariate distribution of the target population can be different from the distribution in the source population. 

\begin{figure}[ht!]
    \caption{An example of a prediction error modifier, $X$.}
        \begin{center}
        \centerline{\includegraphics[width = 3.5in]{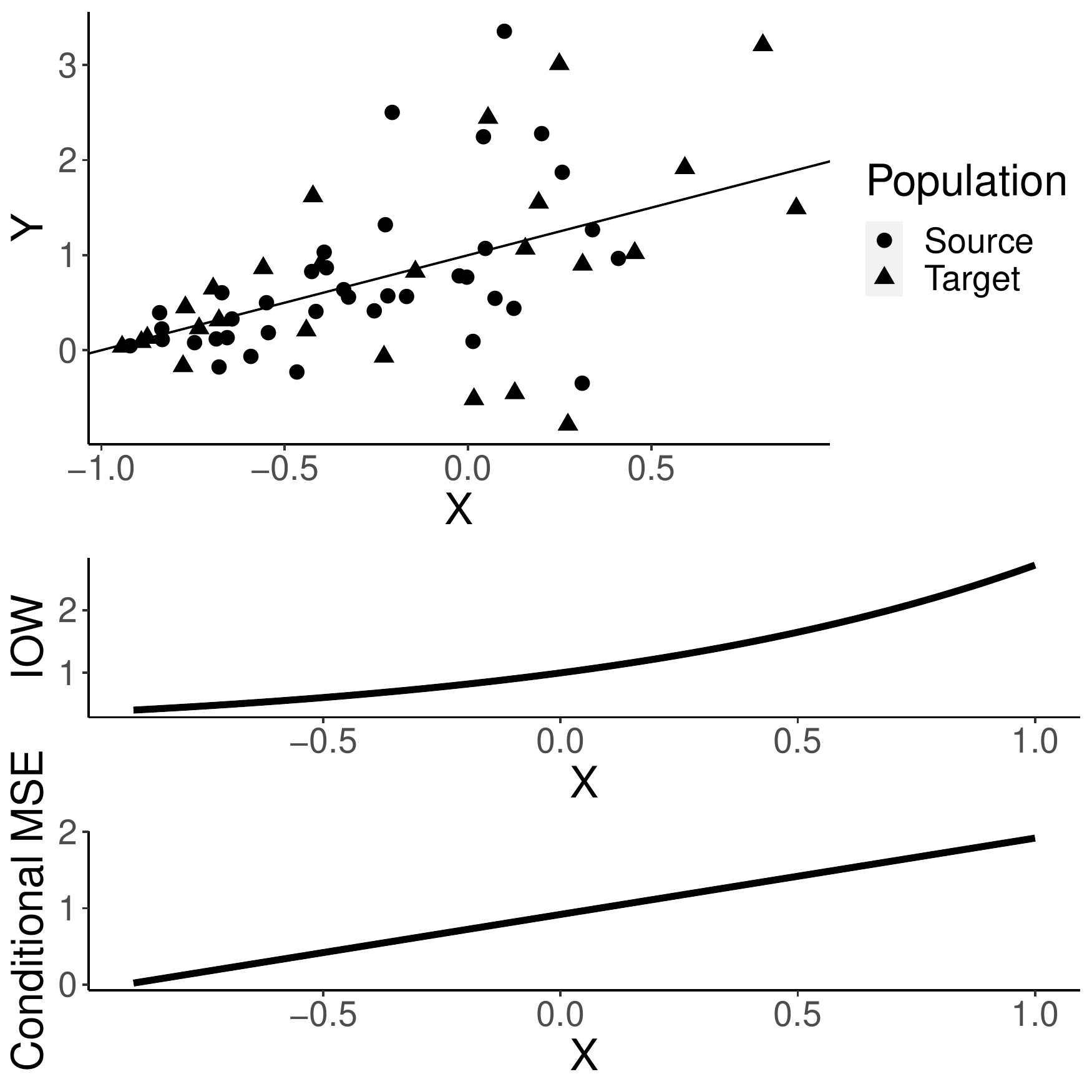}}
    \end{center}
    \caption*{The top panel shows a scatter-plot of the data (including the unobserved target population outcomes) and the solid black line is the true conditional expectation function $\E[Y|X,S=1]$. The middle panel shows the inverse-odds weights (IOW) as a function of $X$ and the bottom panel shows the conditional mean squared error (MSE) as a function of $X$. In these artificial data, larger values of $X$ have higher probability of being from the target population, $S=0$ (corresponding to lower odds of being from the source population and higher inverse-odds weights) and higher MSE. Hence, $X$ is a prediction error modifier that is differentially distributed between the source and the target population. This leads to the source population MSE being smaller than the target population MSE (0.47 versus 0.74).}
    \label{fig:PM}
\end{figure}

Figure \ref{fig:PM} shows an example of a prediction error modifier that is differently distributed between the source and target population resulting in an MSE in the target population that is higher than the MSE in the source population; as the covariate vector in the example is one dimensional $X$ and $Z$ are equal. In the middle panel of Figure \ref{fig:PM} we plot the inverse-odds weights as a function of the prediction error modifier $X$; in the bottom panel we plot the conditional squared errors as a function of $X$. Because both the conditional squared errors and the inverse-odds weights (and therefore the probability of being from the target population) increase as $X$ increases, the target population MSE (which is equal to the expectation of the squared errors) is larger than the source population MSE. Hence, directly using the source population MSE in the context of the target population would lead to over-optimism about model performance.

\subsection*{Assessing model performance in the target population}

In our setup, where outcome information is only available from the sample of the source population, we need to account for differences in the data distribution between the source population and the target population to assess model performance in the target population. 
Proposition \ref{thm2} in Appendix \ref{app:ID} shows that, under the setup described previously and conditions A1 and A2, $\psi_{\widehat \beta}$ is identifiable using source and target population data through the expression 
\begin{equation*}
\psi_{\widehat \beta} = \E[\E[(Y- g_{\widehat \beta}(X))^2|X, S=1, D_{\text{\tiny test}}=1]|S=0, D_{\text{\tiny test}}=1],
\end{equation*}
or equivalently using an inverse-odds weighting expression
\begin{equation}
\label{ipw-d}
\psi_{\widehat \beta} = \frac{1}{\Pr[S=0|D_{\text{\tiny test}}=1]}\E\left[\frac{I(S=1) \Pr[S=0|X,D_{\text{\tiny test}}=1]}{\Pr[S=1|X,D_{\text{\tiny test}}=1]}(Y- g_{\widehat \beta}(X))^2\bigg|D_{\text{\tiny test}}=1\right].
\end{equation}
Here, $D_{\text{\tiny test}}$ is an indicator for whether an observation is in the source or target test data.

The identifiability result in expression \eqref{ipw-d} suggests the following inverse-odds weighting estimator \cite{zadrozny2004learning,shimodaira2000improving} for the target population MSE:
\begin{equation}
\label{IPE-est}
\widehat \psi_{\widehat \beta} = \frac{\sum\limits_{i=1}^n I(S_i=1,D_{\text{\tiny test},i}=1) \widehat o(X_i) \left(Y_i - g_{\widehat \beta}(X_i)\right)^2}{\sum\limits_{i=1}^n I(S_i=0,D_{\text{\tiny test},i}=1)},
\end{equation}
where $\widehat o(X)$ is an estimator for the inverse-odds weights in the test set, $\dfrac{\Pr[S=0|X, D_{\text{\tiny test}}=1]}{\Pr[S=1|X, D_{\text{\tiny test}}=1]}$. To ensure independence between the data used to train the model and the data used to evaluate the model, we propose to use inverse-odds weights estimated using the training set for model building and inverse-odds weights estimated using the test set for estimating model performance.

An important feature of our result is that it does not require the prediction model to be correctly specified, that is, we do not assume that $g_{\widehat \beta}(X)$ converges to the true conditional expectation of the outcome in the source population, $\E[Y|X,S=1]$. This implies that model performance measures in the target population are identifiable and estimable, both for misspecified and correctly specified models. Informally, our identifiability results require the existence of a common underlying model for the source and target population (condition A1), but they do not require the (much less plausible) assumption that investigators can correctly specify that model.


So far we have focused on the scenario where the prediction model is built using the training data and is evaluated using the test data, and where the entire composite dataset (formed by appending data from the source and target population) is split into a test and a training set that are used for model estimation and assessment. In some cases an established model is available (e.g., one developed using external data) and the goal of the analysis is limited to assessing model performance in the target population. In that case, no data from the source or target population need to be used for model development and all available data can be used to evaluate model performance and treated as a part of the ``test set''.

We should note here that provided the prediction model is correctly specified, exchangeability in mean over $S$, that is $\E[Y|X,S=1] = \E[Y|X,S=0]$, is sufficient for the parameter $\beta$ to be identifiable using data from the source population alone. Exchangeability in mean over $S$ is a weaker condition than condition A1; that is, condition A1 implies exchangeability in mean, but the converse is not true. 
Exchangeability in mean, however, is not sufficient for transporting measures of model performance, such as the MSE. In Appendix \ref{App:A1} we give an example of a setting where exchangeability in mean holds but it is not sufficient to identify the target population MSE.

\section*{Model and tuning parameter selection}

Up to now we have proceeded as if the source population data in the training set are used to estimate parameters of a pre-specified parametric model, without employing any form of model selection (e.g., variable choice or other specification search) or tuning parameter selection. Yet, when developing prediction models, analysts often select between multiple different models and statistical learning algorithms usually have one or more tuning parameters. Importantly, data-driven methods for model and tuning parameter selection, such as cross-validation-based procedures, rely on optimizing some measure of model performance, such as the MSE.

Consider, for instance, tuning parameter selection using $K$-fold cross-validation. In such an analysis, we split the data into $K$ mutually exclusive subsets (``folds'') and for each value of the tuning parameter we build the model with the selected tuning parameter value on $K-1$ of the folds and estimate a measure of model performance on the fold that is not used for model building. This process is repeated where each of the $K$ folds is left out of the model building process, resulting in $K$ estimates of model performance. The final cross-validated estimator of model performance associated with the tuning parameter value is the average of the $K$ estimators. The cross-validated value of the tuning parameter is selected as the value of the tuning parameter that optimizes the cross-validated estimator of model performance.

Clearly, data-driven model and tuning parameter selection relies on estimating measures of model performance. Furthermore, tailoring the cross-validated model for use in the target population and tuning parameter selection to improve model performance for use in the target population require incorporating the results from the two preceding sections to account for differences in the distribution of covariates between the source and target population. Specifically, when prediction error modifiers have a different distribution in the source and the target population, cross-validated measures of model performance calculated using the source population data are biased estimators of model performance in the target population. Inverse-odds weighting estimators can adjust for that bias and failing to adjust for this bias when performing cross-validation is likely to lead to sub-optimal model or tuning parameter selection in the context of the target population. 

\section*{Illustration using simulated data}
\label{sec:Ill}

In this section we use simulated data to illustrate (i) the performance of correctly and incorrectly specified prediction models when used with or without inverse-odds of participation weights; (ii) the potential for bias resulting from the naive (unweighted) MSE estimator that uses only the source population data to estimate the target population MSE; and (iii) the ability to adjust for that bias using the inverse-odds weighting estimator.

\paragraph{Data generation:} We simulated the outcome using the linear model $Y = 1 + X + 0.5  X^2 + \varepsilon$, where $\varepsilon \sim \mathcal{N}(0, X)$ and $X \sim Uniform(0,10)$. Under this model, the errors are heteroscedastic because the error variance directly depends on the covariate $X$. We simulated participation in the source data using a logistic regression model $\ln\left(\frac{\Pr[S=1|X]}{1 - \Pr[S=1|X]} \right) = 1.5 - 0.3 X$. We set the total sample size to $1000$ and the source and target population data were randomly split in a 1:1 ratio into a training and a test set.

\begin{figure}[ht!]
    \caption{An example of simulated data used to illustrate transportability of prediction models.}
        \begin{center}
        \centerline{\includegraphics[width = 3in]{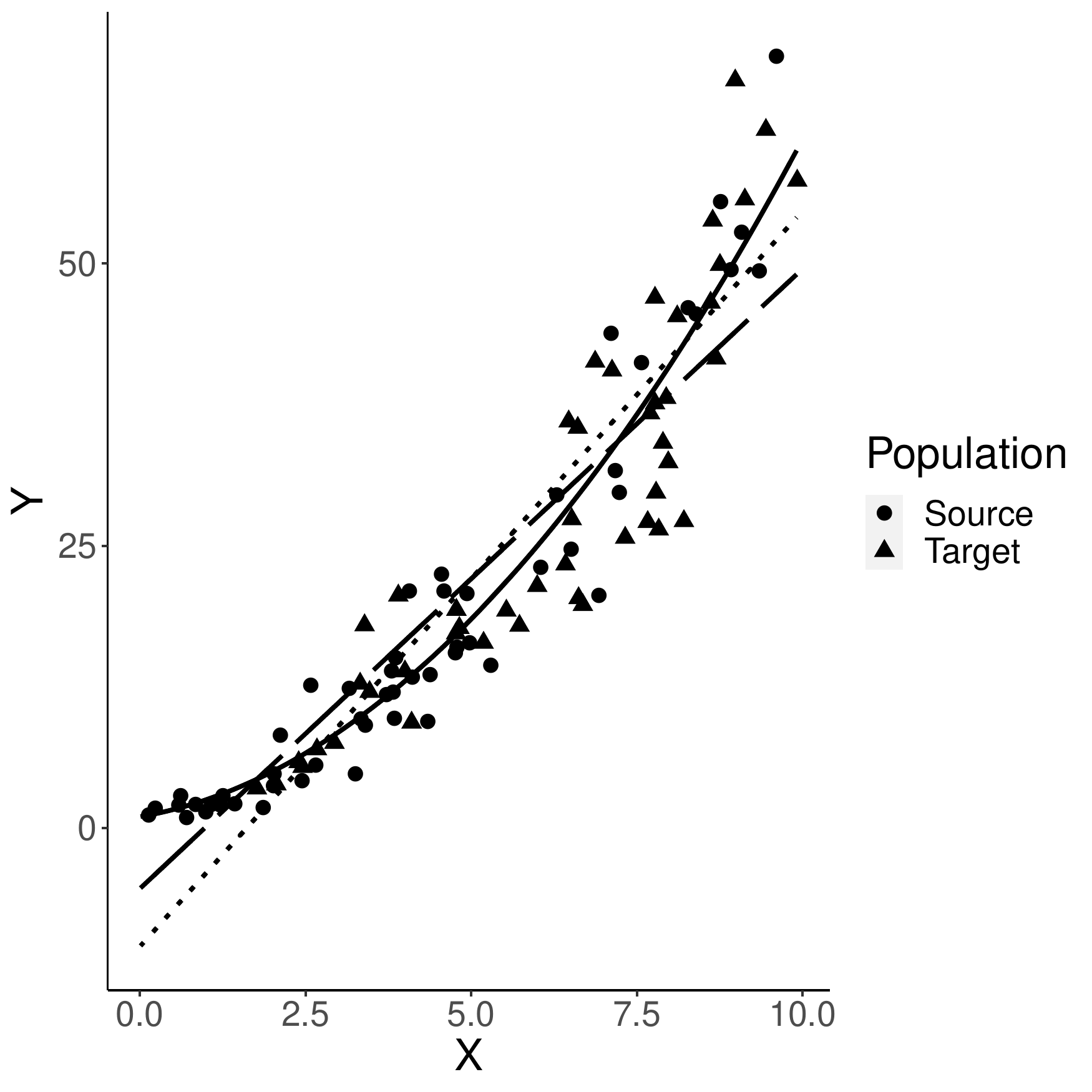}}
    \end{center}
    \caption*{The solid curve is $\E[Y|X, S=1]$, the dashed line is the large-sample limit when estimating the misspecified model without weighting, and the dotted line is the large-sample limit when estimating the misspecified model using inverse-odds weights. The weighted estimation gives more influence to observations with higher values of $X$, compared to unweighted estimation, because higher values of $X$ are associated with higher odds of a sampled observation being from the target population (i.e., lower odds of being from the source population, corresponding to higher inverse-odds weights). This is seen in the figure as for high values of $X$ the weighted model better approximates $\E[Y|X, S=1]$ compared to the unweighted model, but the opposite is true for smaller values of $X$.}
    \label{fig:Sim}
\end{figure}

Under this data generating mechanism, the target population MSE is larger than the source population MSE and both conditions A1 and A2 are satisfied. We considered two prediction models, a correctly specified linear regression model that included main effects of $X$ and $X^2$ and a misspecified linear regression model that only included the main effect of $X$. We also considered two approaches for estimating each posited prediction model: ordinary least squares regression (unweighted, OLS) and weighted least squares regression (WLS) where the weights were equal to the inverse of estimated odds of participation in the source data training set. We estimated the inverse-odds of participation in the training set,  $\Pr[S=0|X, D_{\text{\tiny train}} = 1]/\Pr[S=1|X, D_{\text{\tiny train}}=1]$, using a correctly specified logistic regression model for $\Pr[S=1|X, D_{\text{\tiny train}}=1]$. Figure \ref{fig:Sim} highlights the relationship between the correct model, and the large-sample limits of the weighted and unweighted misspecified models. For the inverse-odds weighting estimator $\widehat \psi_{\widehat \beta}$, we estimated the odds weights $\widehat o(X)$ in the test set by fitting a correctly specified logistic regression model for $\Pr[S=1|X,D_{\text{\tiny test}}=1]$ using the test set data.

\paragraph{Simulation results:}  The results from $10,000$ runs of the simulation are presented in Table \ref{sim:tab}. For both OLS and WLS estimation of the prediction model, the correctly specified model resulted in smaller average target population and source population MSE estimates compared with the misspecified model. When comparing the performance of OLS and WLS estimation of the prediction model in the target population OLS performed slightly better than WLS when the model was correctly specified (average MSE of $45.8$ vs.~$46.2$). When the prediction model was incorrectly specified, OLS performed worse than WLS (average MSE of $66.3$ vs.~$58.0$). The last column in the Table shows that the average of the inverse-odds weighting MSE estimator across the simulations was very close to the true target population MSE (obtained via numerical methods) for all combinations of model specifications and use of weights. In all scenarios of this simulation, the source population MSE estimator was substantially lower than the target population MSE. Hence, using the estimated source population MSE as an estimator for the target population MSE would lead to substantial underestimation of the MSE (i.e.,~showing model performance to be better than it is in the context of the target population). In contrast, the inverse-odds weighting estimator would give an accurate assessment of model performance in the target population.

\section*{Nested designs}

Thus far, we have focused on the non-nested sampling design. Nested sampling designs are an alternative approach where the source population is a subset of the target population of interest \cite{dahabreh2019generalizing, dahabreh2019study, lu2019causal}. Examples of such nested designs arise when the sample from the source population, from which outcome information is available, can be embedded within a larger cohort (e.g., via record linkage techniques) that can be viewed as representing the target population. Our results can be applied, with minor modifications, to nested designs. In Appendix \ref{app:nested}, we prove an identification result for nested designs and provide an estimator for loss-based measure of target population model performance.

\section*{Discussion}

We considered transporting prediction models to a different population than was used for original model development, when outcome and covariate data are available on a simple random sample from the source population and covariate information is available on a simple random sample from the target population. We described the adjustments needed when the covariate distribution differs between the source and target population and provided identification results.
We discussed how to tailor the prediction model to the target population and how to calculate measures of model performance in the context of the target population, without requiring the prediction model to be correctly specified. We also examined tailoring data-driven model and tuning parameter selection to the target population. The key insight is that most measures of model performance average over the covariate distribution and, as a result, estimators of these measures obtained in data from the source population will typically be biased for the corresponding measures in the target population, when the covariate distribution differs between the two populations.



To simplify the exposition, throughout this paper we have assumed that the covariates needed to satisfy the conditional independence condition (A1) are the same as the covariates used in the prediction model. In practice, the set of covariates needed to satisfy condition A1 may be much larger than the set of covariates that are practically useful to include in the prediction model. The identifiability results in our paper can be easily modified to allow for the two sets of covariates to be different.

\clearpage
\bibliographystyle{unsrt}
\bibliography{References}
\noindent

\clearpage
\section*{Figures}

\clearpage 
\section*{Table}

\begin{table}[ht!]
\caption{Target population mean squared error (MSE), average of the source data MSE estimators, and the estimators for the target population MSE that weight observations by the inverse-odds of being from the source population.}\label{sim:tab}
\centering
\renewcommand{\arraystretch}{1.1}
\footnotesize
\begin{tabular}{|l|c|c|c|}
\hline
    \begin{tabular}[c]{@{}l@{}}Model specification, \\ estimation approach\end{tabular}    &  \begin{tabular}[c]{@{}l@{}} True target  \\  population MSE \end{tabular}      & \begin{tabular}[c]{@{}l@{}}Average of unweighted   \\  MSE estimator \end{tabular}    & \begin{tabular}[c]{@{}l@{}}Average of weighted \\  MSE estimator \end{tabular} \\ \hline
\begin{tabular}[c]{@{}l@{}}Correctly specified, \\ OLS\end{tabular}   &  45.8            & 22.5                  & 45.8          \\ \hline
\begin{tabular}[c]{@{}l@{}}Incorrectly specified, \\ OLS\end{tabular} & 66.3             & 34.5                  & 66.3          \\ \hline
\begin{tabular}[c]{@{}l@{}}Correctly specified, \\ WLS\end{tabular}     & 46.2             & 22.8                  & 46.2         \\ \hline
\begin{tabular}[c]{@{}l@{}}Incorrectly specified, \\ WLS\end{tabular} &          58.0   &  43.6                 & 57.9          \\ \hline
\end{tabular}
\caption*{Correctly specified and incorrectly specified refers to the specification of the posited prediction model. OLS = model estimation using ordinary least squares regression (unweighted); WLS =  model estimation using weighted least squares regression with weights equal to the inverse of the odds of being from the source population. Weighted MSE estimator results were obtained using the estimator in equation \eqref{IPE-est}. Results were averaged over $10,000$ simulations. The true target population MSE was obtained using numerical methods.}

\end{table}

\newpage 
\appendix 

\section{Proofs of key results}\label{App:proof}
\renewcommand{\theequation}{A.\arabic{equation}}
\setcounter{equation}{0}

\subsection{Identifiability for non-nested designs}

\subsubsection*{Proof of identifiability of target population MSE} 
\label{app:ID}

We will provide the identifiability result for a general loss function $L(Y,g_{\widehat \beta}(X))$. Many common performance measures, including the mean squared error, absolute error, and the Brier score, are special cases of expected loss functions. We define $D_{\text{ \tiny test}}$ as an indicator if an observation is in the source or target test data.

\begin{proposition}
\label{thm2}
Under conditions A1 and A2 and when the source and target data are obtained by separate simple random sampling of the corresponding underlying populations, with potentially unknown sampling probabilities, then the target population MSE, $\psi_{\widehat \beta}$, is identifiable as
\begin{equation}
\label{g-form-d-app}
\psi_{\widehat \beta} = \E[\E[(Y- g_{\widehat \beta}(X))^2|X, S=1, D_{\emph{\tiny test}}=1]|S=0, D_{\emph{\tiny test}}=1];
\end{equation}
or, using an inverse-odds weighting representation,
\begin{equation}
\label{ipw-d-app}
\psi_{\widehat \beta} = \frac{1}{\Pr[S=0|D_{\emph{\tiny test}}=1]}\E\left[\frac{I(S=1) \Pr[S=0|X,D_{\emph{\tiny test}}=1]}{\Pr[S=1|X,D_{\emph{\tiny test}}=1]}(Y- g_{\widehat \beta}(X))^2\bigg|D_{\emph{\tiny test}}=1\right].
\end{equation}
All quantities in expressions \eqref{g-form-d-app} and \eqref{ipw-d-app} condition on $D_{\emph{\tiny test}}=1$ and can therefore be calculated using the available test data. 
\end{proposition}\textbf{}
\begin{proof}
For the first representation we have
\begin{align*}
\psi_{\widehat \beta} &= \E[L(Y, g_{\widehat \beta}(X))|S=0] \\
&= \E\left[\E[L(Y,g_{\widehat \beta}(X))|X,S=0]|S=0\right] \\
&= \E\left[\int L(y,g_{\widehat \beta}(X)) dF(y|X, S=0)\bigg|S=0\right] \\
&= \E\left[\int L(y,g_{\widehat \beta}(X)) dF(y|X, S=1)\bigg|S=0\right] \\
&= \E\left[\E[L(Y,g_{\widehat \beta}(X))|X,S=1]|S=0\right],
\end{align*}
where the first equality follows from the definition of $\psi_{\widehat \beta}$, the second from the law of iterated expectations, the third from the definition of conditional expectation, and the fourth from identifiability condition A1. All expectations conditional on $(X,S=1)$ in the above formula are well defined by the positivity condition A2. Rewrite
\begin{align*}
\psi_{\widehat \beta} &= \E[\E[L(Y, g_{\widehat \beta}(X))|X, S=1]|S=0] \\
&= \int \E[L(Y- g_{\widehat \beta}(X))|X=x, S=1] dF(x|S=0).
\end{align*}
The conditional expectation $\E[L(Y, g_{\widehat \beta}(X))|X=x, S=1]$ is identifiable because, under the non-nested sampling design, data are available from a random sample of observations from  the source population ($S=1$). Furthermore, the conditional distribution $F(x|S=0)$ is also identifiable because, under the non-nested sampling design, data are available from a random sample of observations from the target population ($S=0$). More formally, the random sampling ensures that 
\[
\psi_{\widehat \beta} = \E[\E[L(Y,g_{\widehat \beta}(X))|X, S=1, D_{\text{\tiny test}}=1]|S=0, D_{\text{\tiny test}}=1].
\]

For the inverse-odds weighting representation
\begin{align*}
\psi_{\widehat \beta} &= \E[\E[L(Y, g_{\widehat \beta}(X))|X, S=1, D_{\text{\tiny test}}=1]|S=0, D_{\text{\tiny test}}=1] \\
&= \E\left[\E\left[\frac{I(S=1)}{\Pr[S=1|X,D_{\text{\tiny test}}=1]}L(Y, g_{\widehat \beta}(X))\bigg|X, D_{\text{\tiny test}}=1\right]\bigg|S=0, D_{\text{\tiny test}}=1\right] \\
&= \frac{1}{\Pr[S=0|D_{\text{\tiny test}}=1]}\E\left[I(S=0)\E\left[\frac{I(S=1)}{\Pr[S=1|X,D_{\text{\tiny test}}=1]}L(Y, g_{\widehat \beta}(X))\bigg|X,D_{\text{\tiny test}}=1\right]\Bigg| D_{\text{\tiny test}}=1\right] \\
&= \frac{1}{\Pr[S=0|D_{\text{\tiny test}}=1]}\E\left[\E\left[\frac{I(S=1) \Pr[S=0|X,D_{\text{\tiny test}}=1]}{\Pr[S=1|X,D_{\text{\tiny test}}=1]}L(Y,g_{\widehat \beta}(X))\bigg|X,D_{\text{\tiny test}}=1\right]\bigg|D_{\text{\tiny test}}=1\right] \\
&= \frac{1}{\Pr[S=0|D_{\text{\tiny test}}=1]}\E\left[\frac{I(S=1) \Pr[S=0|X,D_{\text{\tiny test}}=1]}{\Pr[S=1|X,D_{\text{\tiny test}}=1]}L(Y, g_{\widehat \beta}(X))\bigg|D_{\text{\tiny test}}=1\right].
\end{align*}
For the fourth equality we have used that
\begin{align*}
& \E\left[I(S=0)\E\left[\frac{I(S=1)}{\Pr[S=1|X,D_{\text{\tiny test}}=1]} L(Y, g_{\widehat \beta}(X))\Bigg|X, D_{\text{\tiny test}}=1\right]\Bigg|D_{\text{\tiny test}}=1 \right] \\
&=\E\left[\E\left[I(S=0)\E\left[\frac{I(S=1)}{\Pr[S=1|X,D_{\text{\tiny test}}=1]} L(Y, g_{\widehat \beta}(X))\Bigg|X, D_{\text{\tiny test}}=1\right] \Bigg| X, D_{\text{\tiny test}}=1 \right]\Bigg|D_{\text{\tiny test}}=1 \right] \\
&= \E\left[\E\left[\frac{I(S=1)}{\Pr[S=1|X,D_{\text{\tiny test}}=1]} L(Y, g_{\widehat \beta}(X))\Bigg|X, D_{\text{\tiny test}}=1\right] \E[I(S=0) | X,D_{\text{\tiny test}}=1]\Bigg| D_{\text{\tiny test}}=1 \right] \\
&= \E\left[\E\left[\frac{I(S=1) \Pr[S=0|X,D_{\text{\tiny test}}=1]}{\Pr[S=1|X,D_{\text{\tiny test}}=1]} L(Y, g_{\widehat \beta}(X))\Bigg|X, D_{\text{\tiny test}}=1\right]\Bigg| D_{\text{\tiny test}}=1 \right]
\end{align*}
All of the quantities in
\[
\frac{1}{\Pr[S=0|D_{\text{\tiny test}}=1]}\E\left[\frac{I(S=1) \Pr[S=0|X,D_{\text{\tiny test}}=1]}{\Pr[S=1|X,D_{\text{\tiny test}}=1]}L(Y, g_{\widehat \beta}(X))\bigg|D_{\text{\tiny test}}=1\right].
\]
condition on $D_{\text{\tiny test}}=1$ and are therefore identifiable using the observed data.
\end{proof}
\subsubsection*{Proof of identifiability of inverse-odd weights} 
\label{app:ID2}
Let $D_{\text{\tiny train}}$ be an indicator if data from an observation is in the training set and used to estimate the inverse-odds weights. The sampling design assumes that $\Pr[D_{\text{\tiny train}}=1|X,S=1] =a$ for some potentially unknown constant $a > 0$; and $\Pr[D_{\text{\tiny train}}=1|X,S=0] = b$ for some potentially unknown constant $b >0$. By the random formation of the test and the training set, the inverse-odds weights in the test and the training set are equal. But, to ensure independence between the data used to train the model and the data used to evaluate the model we propose to use inverse-odds weights estimated using the training set for model building and the inverse-odds weights estimated using the test set for estimating model performance.

\subsubsection*{Proof of expression \ref{OR-ID} from the main text}
Recall that the sampling design assumes that $\Pr[D_{\text{\tiny train}}=1|X,S=1] =a$ for some potentially unknown constant $a > 0$ and $\Pr[D_{\text{\tiny train}}=1|X,S=0] = b$ for some potentially unknown constant $b >0$. Using that, we have
\begin{align*}
\frac{\Pr[S=0|X,D_{\text{\tiny train}}=1]}{\Pr[S=1|X,D_{\text{\tiny train}}=1]} &= \frac{\Pr[S=0,D_{\text{\tiny train}}=1|X]}{\Pr[S=1, D_{\text{\tiny train}}=1|X]} \\
&=  \frac{\Pr[S=0|X]}{\Pr[S=1|X] } \times \frac{\Pr[D_{\text{\tiny train}}=1|X, S=0]}{\Pr[D_{\text{\tiny train}}=1|X,S=1]} \\
&= \frac{\Pr[S=0|X]}{\Pr[S=1|X] } \times \frac{\Pr[D_{\text{\tiny train}}=1| S=0]}{\Pr[D_{\text{\tiny train}}=1|S=1]} \\
&= \frac{\Pr[S=0|X]}{\Pr[S=1|X] } \times \frac{b}{a} \\
&\propto \frac{\Pr[S=0|X]}{\Pr[S=1|X] }.
\end{align*} \qed

\section{Identification and estimation in nested designs}
\label{app:nested}

Consider a nested design where the source population is a subset of a larger target population of interest. We assume that covariate data, $X$, are available on all target population members, but outcome data, $Y$, are only available on everyone in the source population. The data is assumed to be realizations of
\[
\{(X_i, S_i, S_i \times Y_i, i = 1, \ldots, n \},
\]
where $n$ is the total number of observations (i.e., the total number of individuals in a cohort representing the target population and in which the sample from the source population is nested) and $S$ is the indicator of an observation coming from the source population ($S=1$ for observations in the source population and $S=0$ for observations not in the source population).

For nested designs the target parameter is defined as
\[
\phi_{\widehat \beta} = \E\left[L(Y,g_{\widehat \beta}(X))\right].
\]
We introduce the following modified identifiability conditions:
\begin{enumerate}
\item[B1.] For every $x$ such that $f(X = x) \neq 0$,
\[
f(Y|X=x, S=1) = f(Y|X=x).
\]
\item[B2.] For every $x$ such that $f(X=x) \neq 0$, $\Pr[S = 1 | X = x] > 0$ .
\end{enumerate}
\begin{proposition}\label{thm3}
Under conditions B1 and B2, $\phi_{\widehat \beta}$ can be written as the observed data functional 
\begin{equation}
\label{out-nest}
\phi_{\widehat \beta} =  \E\left[\E\left[L(Y,g_{\widehat \beta}(X))\big|X, S=1, D_{\text{\tiny test}}=1\right]D_{\text{\tiny test}}=1\right].
\end{equation}
Or using the inverse probability weighting representation
\begin{equation}
\label{ipw-nest}
\phi_{\widehat \beta} = \E\left[\frac{I(S=1)}{\Pr[S=1|X, D_{\text{\tiny test}}=1]}L(Y,g_{\widehat \beta}(X))\Bigg|  D_{\text{\tiny test}}=1\right]. 
\end{equation}
\end{proposition}

\subsubsection*{Proof of Proposition \ref{thm3}:} 

We have 
\begin{align*}
\phi_{\widehat \beta} &= \E\left[L(Y,g_{\widehat \beta}(X))\right] \\
& = \E\left[\E\left[L(Y,g_{\widehat \beta}(X))\big|X\right]\right] \\
& = \E\left[\E\left[L(Y,g_{\widehat \beta}(X))\big|X, S=1\right]\right]\\
& = \E\left[\E\left[L(Y,g_{\widehat \beta}(X))\big|X, S=1, D_{\text{\tiny test}}=1\right]\Big| D_{\text{\tiny test}}=1\right].
\end{align*}
For the inverse probability weighting representation
\begin{align*}
\phi_{\widehat \beta} &= \E\left[\E\left[L(Y,g_{\widehat \beta}(X))\big|X, S=1, D_{\text{\tiny test}}=1\right]\Bigg|D_{\text{\tiny test}} = 1\right] \\
&= \E\left[\E\left[\frac{I(S=1)}{\Pr[S=1|X,D_{\text{\tiny test}}=1]}L(Y,g_{\widehat \beta}(X))\bigg|X, D_{\text{\tiny test}}=1\right]\Bigg| D_{\text{\tiny test}}=1 \right] \\
&= \E\left[\frac{I(S=1)}{\Pr[S=1|X,D_{\text{\tiny test}}=1]}L(Y,g_{\widehat \beta}(X))\Bigg | D_{\text{\tiny test}}=1\right] \\
&=  \frac{1}{\Pr[D_{\text{\tiny test}}=1]} \E\left[\frac{I(S=1,D_{\text{\tiny test}}=1 )}{\Pr[S=1|X,D_{\text{\tiny test}}=1]}L(Y,g_{\widehat \beta}(X)) \right],
\end{align*}
which establishes the identifiability of $\phi_{\widehat \beta}$. \qed

Using plug-in estimators into identifiability expression \eqref{ipw-nest} gives the inverse probability weighting estimator for nested designs. That is, 
\[
\widehat \phi_{\widehat \beta} = \frac{\sum_{i=1}^n \frac{I(S_i=1, D_{\text{\tiny test},i} =1)}{\widehat p(X_i)}L(Y_i,g_{\widehat \beta}(X_i))}{\sum_{i=1}^n I(D_{\text{\tiny test},i} =1)} ,
\]
where $\widehat p(X)$ is an estimator for $\Pr[S=1|X,D_{\text{\tiny test}}=1]$.

\clearpage
\section{Inverse-odds weighting estimators can be biased under mean exchangeability}\label{App:A1}
\renewcommand{\theequation}{B.\arabic{equation}}
\setcounter{equation}{0}

\begin{figure}[htbp]
        \begin{center}
        \centerline{\includegraphics[width = 3in]{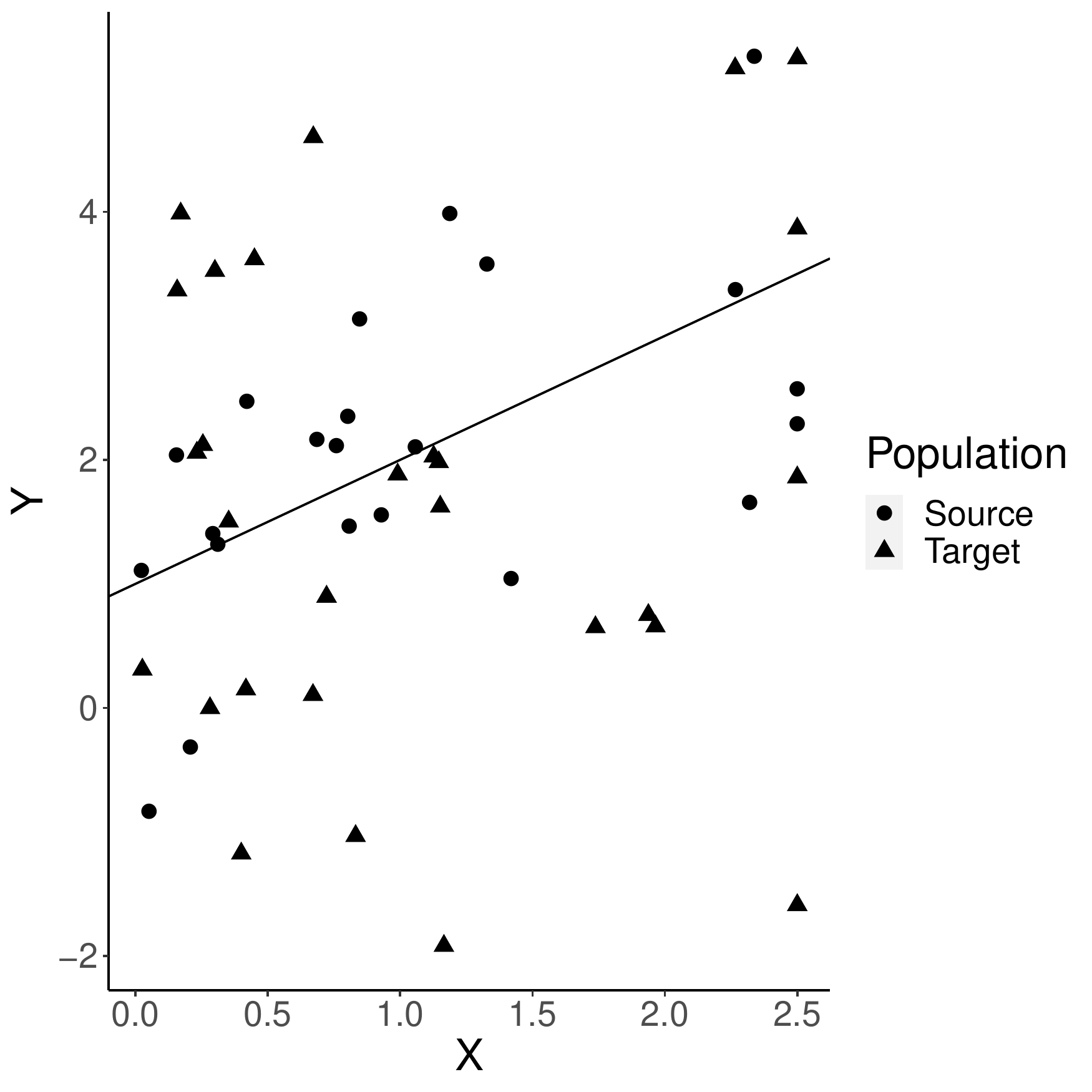}}
    \end{center}
    \caption{An example of a setting where condition A1 does not hold. Here, $\E[Y|X,S=1] = \E[Y|X,S=0]$ (the black line is the true conditional mean for both populations), but $\mbox{Var}[Y|X,S=1] < \mbox{Var}[Y|X,S=0]$ for all values of $X$. In this case, estimators of model performance measures that use weights equal to the inverse-odds of being from the source population (e.g., the MSE estimator in the main text of the paper) will be biased.}
    \label{fig1:PM}
\end{figure}

For correctly specified prediction models, exchangeability in mean over $S$, that is $\E[Y|X,S=1] = \E[Y|X,S=0]$, is sufficient for the parameter $\beta$ to be identifiable using data from the source population alone. Exchangeability in mean over $S$ is a weaker condition than condition A1; that is, condition A1 implies exchangeability in mean, but the converse is not true. Exchangeability in mean, however, is insufficient for transportability of the MSE. This can be seen in Figure \ref{fig1:PM} where $\E[Y|X,S=1] = \E[Y|X,S=0]$ but $\mbox{Var}[Y|X,S=1] \neq \mbox{Var}[Y|X,S=0]$ (and thus assumption A1 does not hold). As the conditional variance is different between the two populations, standardizing to the target population covariate distribution is not sufficient to transport the MSE to the target population.

If the outcome is binary, condition A1 can be written as $\Pr[Y = 1 |X=x, S=1] = \Pr[Y = 1|X=x, S=0]$, so for binary outcomes distributional independence over $S$ is equivalent to exchangeability in mean over $S$. 

\end{document}